\documentclass[8pt,a4]{article}
\usepackage{graphicx}
\usepackage{amssymb,amscd,amsmath}
\usepackage{wrapfig}
\usepackage{feynmf}
\usepackage{appendix}
\usepackage{slashed}
\usepackage{indentfirst}
\usepackage{wrapfig}
\usepackage{appendix}
\usepackage[usenames]{color}
\usepackage{euscript}

\makeatletter \@addtoreset{equation}{section} \makeatother

\newcommand{\absvec}[1]{\left| \mathbf{ #1 } \right|}

\begin{document}

\renewcommand{\abstractname}{Abstract}
\renewcommand{\refname}{References}

\title{\bf ON COLLECTIVE PROPERTIES OF DENSE QCD MATTER}

\author{Igor Dremin, Martin Kirakosyan, Andrei Leonidov\footnote{Also at the Institute of Experimental and Theoretical Physics and Moscow Institute
of Physics and Technology, Moscow, Russia}
\\
\\
\small{\em P.N. Lebedev Physical Institute, Leninsky pr. 53, 119991 Moscow, Russia} }

\date{}
\maketitle

\begin{abstract}
A short review of the two recently analyzed collective effects in dense non-Abelian matter, the photon and dilepton production in nonequilibrium
glasma and polarization properties of turbulent Abelian and non-Abelian plasmas, is given.
\end{abstract}

\bigskip

\bigskip
\newpage
\section{Introduction}

Working out a quantitative description of the properties of dense strongly interacting matter produced in ultrarelativistic heavy ion collisions
presents one of the most fascinating problems in high energy physics. The main goal of the present review is to expand an analysis of various
properties of dense non-Abelian matter presented in \cite{DL10} by discussing several new topics enriching our understanding of the early stages
of ultrarelativistic heavy ion collisions.

The conceptually simplest way of organizing  experimental information obtained at RHIC \cite{hydroRHIC} and LHC \cite{hydroLHC} into a more or
less coherent framework is to describe the late stages of these collisions in terms of standard relativistic hydrodynamical expansion of
primordial quark-gluon matter that, after a short transient period, reaches sufficient level of local isotropization and equilibration allowing
the usage of hydrodynamics in its standard form. In particular, the presence of strong elliptic flow suggest the picture of strongly coupled and,
therefore, low viscosity matter. A detailed discussion of the corresponding issues can be found in, e.g., reviews \cite{H05} and \cite{MSW12}
devoted to RHIC and LHC results respectively.

The actual physical picture is, most likely, much more complicated. In the absence of realistic mechanisms leading to extremely fast
isotropization needed for describing the experimental data within the framework of standard hydro \cite{GJSTV13}, the recent discussion
\cite{ahydro,janik} focused on building a generalization of hydrodynamical approach on systems with anisotropic pressure that naturally arise in
the glasma-based description of the physics of the early stages of heavy ion collisions \cite{KMW95,LM06,DL10}. Of particular interest are results
of \cite{janik} showing, within the AdS-CFT duality paradigm< that hydrodynamic description can be valid for unexpectedly large pressure
gradients. The new paradigm of anisotropic hydrodynamics is, in our opinion, one of the most promising new approaches to the physics of high
energy nuclear collisions.

At the most fundamental level a description of early stages of high energy nuclear collisions in the weak coupling regime is based on the idea
that large gluon density and, correspondingly, large occupation numbers of low energy gluon modes make it natural to use tree-level Yang-Mills
equations with sources in the strong field regime as a major building block for the theoretical description of ultrarelativistic nuclear
collisions. At the early stage a strongly nonisotropic tree-level gluon field configuration arising immediately after collision, the glasma
\cite{KMW95,LM06}, is formed. A very recent development \cite{Blaizot2012} suggests that temporal evolution of glasma involves formation of a
transient coherent object, the gluon condensate. This can have interesting experimental consequences, in particular for photon and dilepton
production \cite{Chiu2012}.

The glasma is, however, is unstable with respect to boost-noninvariant quantum fluctuations \cite{RV06}. At later stages of its evolution these
instabilities were shown to drive a system towards a state characterized by the turbulent Kolmogorov momentum spectrum of its modes \cite{FG12}.
The same Kolmogorov spectrum was earlier discovered in a simplified scalar model of multiparticle production in heavy ion collisions
\cite{DEGV11,EG11}.  A possible relation between these instabilities and low effective viscosity in expanding geometry was recently discussed in
\cite{DEGV12}.

The origin of the initial glasma instabilities and the physical picture underlying the turbulent-like glasma at later stages of its evolution,
however, do still remain unclear. The usual references are here to the Weibel-type instabilities of soft field modes present both in QED and QCD
plasma and having their origin in the momentum anisotropy of hard sources  \cite{W59,M88,PS88,M93,M97,ALM03,ALMY05} and the resulting turbulent
Kolmogorov cascade \cite{AM06a,AM06b}, see also the review \cite{R09} and the recent related development in \cite{KM11,KM12,IRS11,CR11}.

Of major importance to the physics of turbulent quantum field theory that provide another important benchmark for the physics of heavy ion
collisions are also the fixed-box studies in the framework of classical statistical lattice gauge theory \cite{BSS07,BGSS09,BSS09} and a study of
the turbulent cascade in the isotropic QCD matter in \cite{MSW07}. Let us also note that there is no doubt that the genuinely stochastic nature of
the classical Yang-Mills equation \cite{KMOSTY10} should by itself play an important role in the physics of turbulent non-Abelian matter. The
precise relation is however still to be studied.

The importance of turbulent effects makes it natural to study their effects on physically important quantities like shear viscosity. The
corresponding calculation was made in \cite{ABM06,ABM07,ABM11} in a setting generalizing the one used in the earlier studies of turbulent QED
plasma \cite{Tsit,I91}, in which turbulent plasma is described as a system of hard thermal modes and the stochastic turbulent fields characterized
by some spatial and temporal correlation lengths. It was shown that plasma turbulence can serve as a natural source of the above-mentioned
anomalous smallness of viscosity of strongly interacting matter created in high energy heavy ion collisions.

The physics of turbulence, both in liquids \cite{OYS07,ZSIG07,ZSIG08,ZS10,ZSI10,ZS12} and plasma \cite{K02}, is essentially that of space-time
structures that appear at the event-by-event level and, after averaging, give rise to Kolmogorov scaling of the structure functions. The
event-by-event stochastic inhomogeneity of turbulent plasma can therefore play an important role in forming its physical properties. In the
present paper we discuss the turbulent contributions to the most fundamental physical characteristics of plasma, the properties of its collective
modes, plasmons. For simplicity we shall restrict ourselves to considering an Abelian case; the corresponding non-Abelian generalization will
appear in a separate publication \cite{KLM13}. The effects in question can broadly be described as nonlinear Landau damping \cite{I91}. One of the
most interesting effects we see is a nonlinear Landau instability for transverse plasmons at large turbulent fields, i.e. a phenomenon equivalent
to nonlinear Landau damping, but with an opposite sign of the corresponding imaginary part of the response tensor. The origin of the phenomena
considered in the paper is in the stochastic inhomogeneity of the turbulent electromagnetic fields in QED plasma; in this respect they are similar
to the phenomenon of the stochastic transition radiation \cite{T72,KL08a,KL08b}. In particular, similarly to the stochastic transition radiation,
the turbulent contributions to plasmon properties discussed in this paper vanish in the limit of vanishing correlation length of the stochastic
turbulent fields.

\section{Photons and dileptons from glasma}

In this paragraph we shall present, following \cite{Chiu2012}, the main results on photon and dilepton  production in glasma.

\subsection{Thermalizing glasma: basic facts}

Let us first discuss the kinetic framework for glasma evolution as developed in \cite{Blaizot2012}. Evolution of primordial non-Abelian matter
produced in ultrarelativistic heavy ion collisions proceeds through several stages. A natural separation of scales is provided by the saturation
momentum $Q_{sat}$. At earliest times $0 \le t \sim 1/Q_{sat}$ the system can be described in terms of coherent chromoelectric and chromomagnetic
flux tubes. The physics of this stage is that of gluons, so in terms of electromagnetic signals this stage is of no special interest. The density
of quarks becomes substantial at times closer to the thermalization time $t_{therm}$, so in studying the nonequilibrium contributions to photon
and dilepton production we shall focus at the time interval $1/Q_{sat} << t << t_{therm}$.

Let us assume that the gluon momentum density can be written in the following form:
\begin{equation}
  f_g = {\Lambda_s \over {\alpha_s p}} F_g(p/\Lambda),
\end{equation}
where the infrared and ultraviolet momentum cutoffs $\Lambda_s$  and $\Lambda$ are defined as follows. Initially, $\Lambda(t_0) =
\Lambda_s(t_0)\sim Q_{sat}$, whereas at thermalization time $\Lambda_s(t_{therm}) \sim \alpha_s\, T_i$ and $\Lambda(t_{therm}) \sim T_i$. In
estimating the physical cross sections one can use the following simple parametrization of $f_g$:
\begin{eqnarray}\label{exfg}
f_g (E_g) & = & {\rm const.}, \;\;\;\;\;\;\;\;\;\;\; E_g<\Lambda_s \nonumber \\
f_g (E_g) & = & {\rm const.} \frac{\Lambda_s}{E_g}, \;\;\;\;\;\; \Lambda_s<E_g<\Lambda \nonumber \\
f_g (E_g) & = & 0 , \;\;\;\;\;\;\;\;\;\;\;\;\;\;\;\;\;\;\; E_g>\Lambda
\end{eqnarray}

A key physical point of primary importance for the physics of the early stage of heavy ion collisions is that  the phase space for the gluons is
initially over-occupied so that the number density of gluons $n_g$ and their energy density $\epsilon_g$ are related by
\begin{equation}
  n_g/\epsilon_g^{3/4} \sim 1/\alpha_s^{1/4},
\end{equation}
while for a thermally equilibrated Bose system it is necessary that this ratio should be less than a number of the order 1. This fact is a basis
for the hypothesis \cite{Blaizot2012} that the "extra" gluonic degrees of freedom are hidden in the highly coherent color singlet and spin singlet
configuration that can be (approximately) described as a transient Bose condensate with a density
\begin{equation}
    f_{cond} = n_{cond} \delta^{3} (p)
\end{equation}
It is natural to think that the condensate is formed by gluons with masses of order of the natural infrared cutoff of the problem, the Debye mass
\begin{equation}
   M^2_{Debye} \sim \Lambda \Lambda_s
\end{equation}

One of the key features arising in many problems related to physical properties of primordial strongly interacting matter in heavy ion collisions
is the natural asymmetry between longitudinal and transverse degrees of freedom which in the problem under consideration is parametrized by the
fixed asymmetry between the typical transverse and longitudinal pressures $\delta$
\begin{equation} \label{eqn_PL}
  P_L = \delta \, \epsilon
\end{equation}
where $0 \le \delta \le 1/3$, with $\delta=0$ and $\delta=1/3$ corresponding to the free-streaming (thus maximal anisotropy between the
longitudinal and transverse pressure) and the isotropic expansion, respectively. The pressure anisotropy does of course reflect the difference of
characteristic scales of transverse and longitudinal momenta.

The time evolution of the scales $\Lambda_s$ and $\Lambda$ were found to be \cite{Blaizot2012}
\begin{eqnarray}
\Lambda_s & \sim & Q_s \left( {t_0 \over t} \right)^{(4+\delta)/7} \\
 \Lambda & \sim & Q_s \left( {t_0 \over t} \right)^{(1+2\delta)/7}
\end{eqnarray}
which, in turn, leads to the following temporal evolution of the gluon density and Debye mass
\begin{eqnarray}
\Lambda & \sim & Q_s \left( {t_0 \over t} \right)^{(1+2\delta)/7} \\
 M^2_{Debye}  & \sim & Q_{sat}^2  \left( {t_0 \over t} \right)^{(5+3\delta)/7}
\end{eqnarray}
and, finally, the thermalization time:
\begin{equation}
  t_{therm} \sim t_0 \left( 1 \over \alpha_s \right)^{7/(3-\delta)}
\end{equation}
The description of the model is completed by introducing the quark distribution function
\begin{equation}
    f_q =  F_q(p/\Lambda)
\end{equation}
and assuming the proportionality between the condensate density and that of gluons
\begin{equation}
  n_{cond} = \kappa\, n_{gluon}
\end{equation}
where $\kappa$ is a constant of order 1.

\subsection{Electromagnetic Particle Production from the Glasma}

Let us start with deriving a rate of photon production from glasma. The standard expression for the fixed-box photon production rate from the
Compton channel $gq \to \gamma q$ reads:
\begin{equation}
E\dfrac{dN}{d^4x d^3p} \propto F_q (E/\Lambda) \frac{1}{E} \int_{\mu^2}^\infty ds \; (s - \mu^2) \; \sigma_{gq \to \gamma q}(s) \;
\int_{s/4E}^\infty dE_g f_g (E_g) \left[ 1 - F_q (E_g/\Lambda) \right] \label{photfixbox},
\end{equation}
where the lower limit for integration over gluon energy $E_g$ follows from kinematics, $\mu^2$ is an infrared cutoff needed to regularize the
$t(u)$ - channel singularity for diagrams with massless particle exchange which in our case is the Debye mass $\mu^2=\Lambda \Lambda_s$ and
$\sigma_{gq \to \gamma q} (s)$ is the cross-section for gluon Compton effect $gq \to \gamma q$.

In the high energy limit and for small quark densities $F_q$ Eq. (\ref{photfixbox}) simplifies to
\begin{equation}
E\dfrac{dN}{d^4x d^3p} \propto F_q (E/\Lambda) \frac{\Lambda_s \Lambda}{E} \int_{1}^\infty dy \; \ln y \; \int_{\frac{y \Lambda_s
\Lambda}{4E}}^\infty dE_g f_g (E_g)  \label{photfixbox1}
\end{equation}
Using the explicit parametrization of gluon density Eq.~(\ref{exfg}) it straightforward to obtain
\begin{equation}
E\dfrac{dN}{d^4x d^3p} \propto F_q (E/\Lambda) \Lambda \Lambda_s \phi(E/\Lambda), \label{photfixbox4}
\end{equation}
where $\phi(E/\Lambda)$ is some analytically calculable function.  Let us thus assumethat the photon production rate from glasma is given by
\begin{equation}
  {{dN} \over {d^4x dy d^2k_T}} = {\alpha \over \pi} \Lambda_s \Lambda  g(E/\Lambda) \label{photratefb}
\end{equation}

To obtain the overall rate, we need to integrate over longitudinal coordinates.  We assume that the early time expansion is purely longitudinal,
and that in the integration the space-time rapidity is strongly correlated with that of the momentum space-rapidity.  We then have that
\begin{equation}
{{dN} \over {d^2r_T dyd^2k_T}} \sim \alpha \int tdt \Lambda_s \Lambda g(k_T/\Lambda)  \label{rate}
\end{equation}

Using the result of the previous section for the time dependence of the scales $\Lambda$ and $\Lambda_s$, we have
\begin{equation}
   t dt = \kappa^{\prime}~  {{d\Lambda} \over \Lambda}~ {1 \over Q_{sat}^2} \left(Q_{sat} \over \Lambda
   \right)^{14/(1+2\delta)}
\end{equation}
The constant $\kappa^{\prime}$ is of order 1.\\

Doing the integration over $\Lambda$ in Eqn.(\ref{rate}), we find that
\begin{equation}
   {{dN} \over {d^2r_T dy d^2k_T}} \sim \alpha \left(Q_{sat} \over k_T \right)^{\frac{9-3\delta}{1+2\delta}}
\end{equation}

\begin{figure}[tb]
\vspace{0.0cm}
\begin{center}
\scalebox{1.0}[1.0] { \hspace{0.2cm}
  \includegraphics[scale=.88]{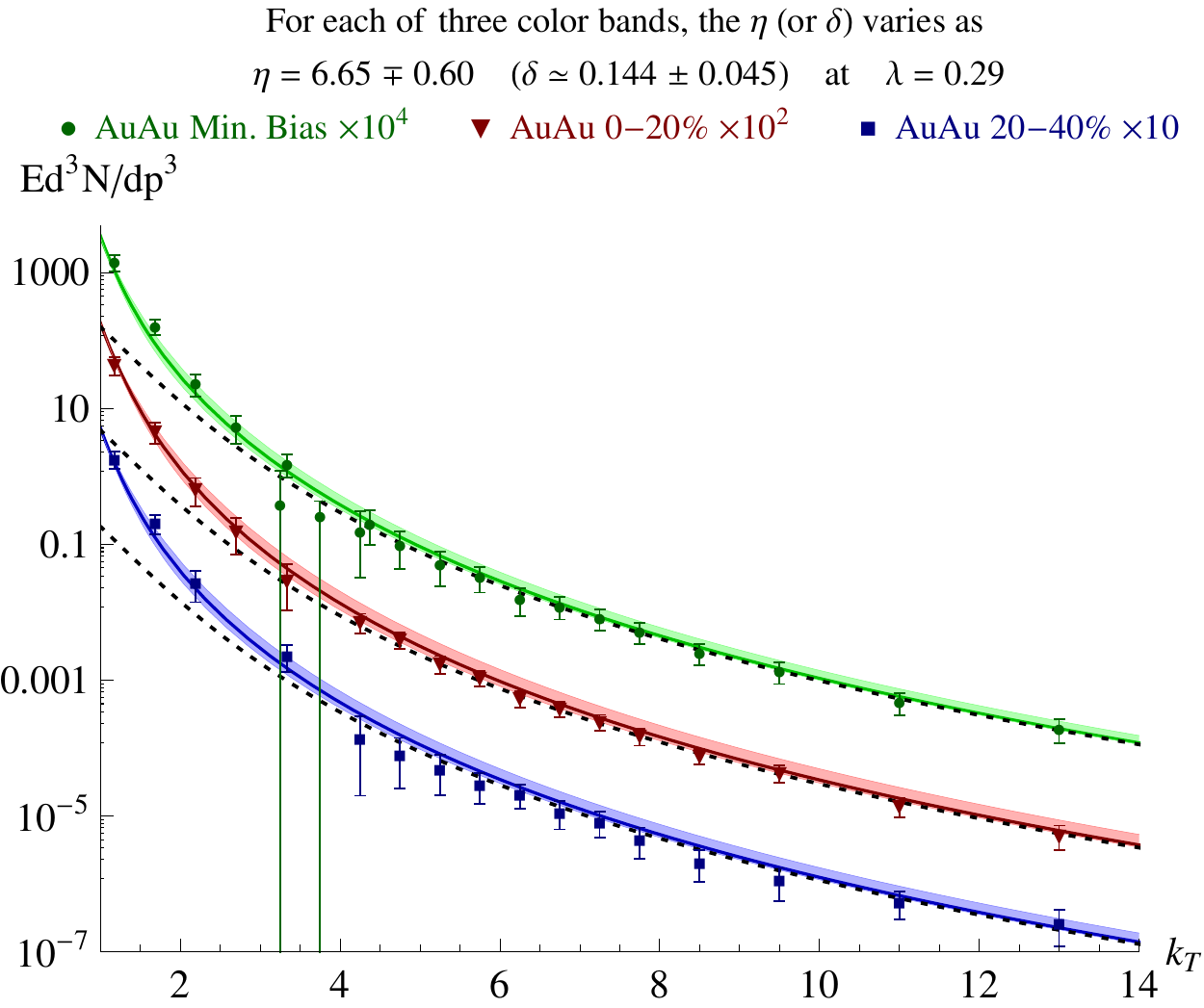} }
\end{center}
\caption{Comparison between the PHENIX photon data and the present model for three centrality bins \cite{Chiu2012}.} \label{photons}
\vspace{0.2cm}
\end{figure}

Now integrating over $d^2r_T$,  and identifying the overlap cross section as proportional to the number of participants, we finally obtain
\begin{equation}
       {{dN_{\gamma}} \over {dy d^2k_T}} =  ~\alpha ~R_0^2 ~N_{part}^{2/3}   \left(Q_{sat} \over k_T \right)^\eta
\end{equation}
where $\eta=(9-3\delta)/(1+2\delta)$. The factor of $N_{part}^{2/3}$ arises because the number of participants in a collision proportional to the
nuclear volume $R^3 \sim N_{part}$ and $R_0$ is a constant with dimensions of a length. The analytical results are compared to RHIC data in
Fig.~\ref{photons}.

From Fig.~\ref{photons} we see that our simple model provides a good description of the experimental data.

The analysis of dilepton production is more complicated  because there are two sources of dileptons.  The first is due to annihilation of quarks
in the Glasma. The expression for the static rate of production of dilepton pairs with invariant mass $M$ for massless quarks and leptons reads
\begin{eqnarray}
\dfrac{dN^{l^+l^-}}{d^4 x d M^2} & \sim & M^2 \sigma_{q{\hat q} \to l^+l^-} (M^2) \int_0^\infty dE_q F_q (E_q/\Lambda) \int_{M^2/4E_q}^\infty
dE_{\bar q} F_{\bar q} (E_{\bar q}/\Lambda)
\nonumber \\
& = & M^2 \sigma_{q{\hat q} \to l^+l^-} (M^2)\, \Lambda^2 \int_0^\infty dy F_q(y) \int_{M^2/4 \Lambda^2 y} dx F_q(x) \label{dil1}
\end{eqnarray}
Taking into account that $M^2 \sigma_{q{\hat q} \to l^+l^-} (M^2) \sim \;{\rm const.}$, we can already see the scaling behavior of the static
dilepton production rate (\ref{dilratefb}), i.e.
\begin{eqnarray}
\dfrac{dN^{l^+l^-}}{d^4 x d M^2} & \sim & \Lambda^2 \, \Phi(M/\Lambda)
\end{eqnarray}
with $\Phi(M/\Lambda)\equiv  \int_0^\infty dy F_q(y) \int_{(M/\Lambda)^2/4y} dx F_q(x)$. To further explicitly demonstrate  the scaling behavior
of the static dilepton production rate (\ref{dilratefb}), let us consider two examples with explicit forms of quark distribution function.

First let us consider a simple hard-cutoff quark distribution function $F_q=\theta(\Lambda - E)$. In this case we can easily obtain
\begin{eqnarray}
\dfrac{dN^{l^+l^-}}{d^4 x d M^2} & \sim & \Lambda^2 \left[ 1-\frac{(M/\Lambda)^2}{4} + \frac{(M/\Lambda)^2}{4} \ln \frac{(M/\Lambda)^2}{4}
\right]
\end{eqnarray}

Second let us consider an exponential quark distribution function $F_q=\exp (-E/\Lambda)$. In this case  we get the following result
\begin{eqnarray}
\dfrac{dN^{l^+l^-}}{d^4 x d M^2} & \sim & \Lambda^2 \, \sqrt{\dfrac{M^2}{\Lambda^2}}  K_1 \left( \dfrac{M}{\Lambda} \right) \nonumber \\
& = & M \Lambda K_1 \left( \dfrac{M}{\Lambda} \right) \equiv \Lambda^2 \, \left[ \dfrac{M}{\Lambda} K_1 \left( \dfrac{M}{\Lambda} \right) \right]
\label{dil2}
\end{eqnarray}
where $K_1 \left( \dfrac{M}{\Lambda} \right)$ is a Bessel function.  This leads to the simple conjecture for the dilepton rate due to the
annihilation mechanism
\begin{equation}
  {{dN_{DY}} \over {d^4x dM^2}} = \alpha^2 \Lambda^2 g^{\prime}(M/\Lambda) \label{dilratefb}
\end{equation}
which, in the direct analogy with the above-described calculation for photons, leads to
\begin{equation} \label{eqn_pair}
{{dN_{DY}} \over {dy dM^2}} \sim \alpha^2 R_0^{\prime 2} ~N_{part}^{2/3} \left(Q_{sat} \over M \right)^\eta
\end{equation}
 with $\eta=4(3-\delta)/(1+2\delta)$.

\begin{figure}[tb]
\vspace{0.0cm}
\begin{center}
\scalebox{1.0}[1.0] { \hspace{0.2cm}
  \includegraphics[scale=.55]{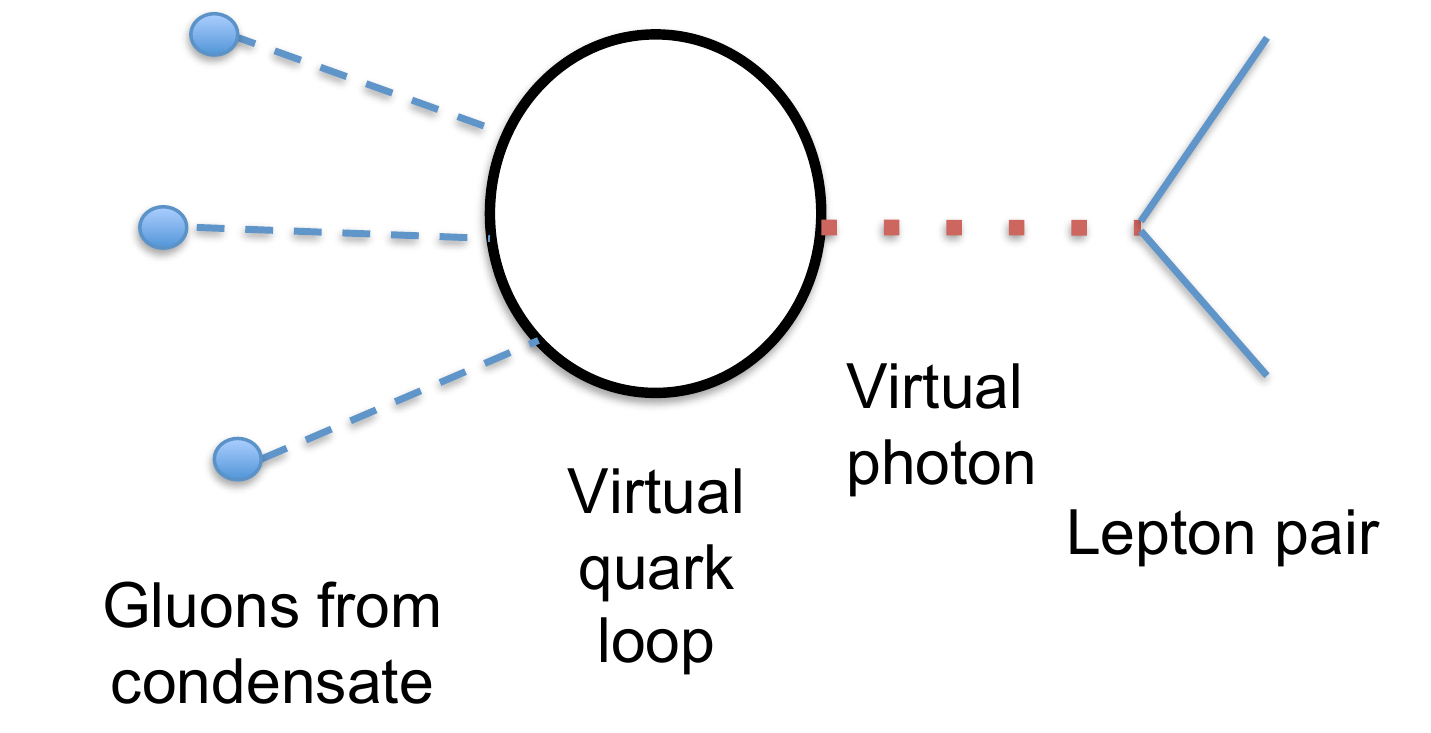} }
\end{center}
\caption{Three gluons from the condensate annihilate into a virtual quark loop, that subsequently decays into a virtual photon and then into a
dilepton. } \label{fig:threegluoni} \vspace{0.2cm}
\end{figure}

The second possible source of dileptons is the annihilation of gluons into a quark loop from which the quarks then subsequently decay into a
virtual photon and eventually the dilepton: see the illustration in the Fig.\ref{fig:threegluoni}.  Such a virtual process is naively suppressed
by factors of $\alpha_s$.  Here however, the gluons arise from a highly coherent condensate, and the corresponding factors of $\alpha_s$ are
compensated by inverse factors $1/\alpha_s$ from the coherence of the condensate.   In other words, the usual power counting for diagrams in terms
of $\alpha_s$ has to be changed when the coherent condensate with high occupation is present.

Here we estimate the rate for the three-gluon decay of the condensate into a dilepton.  On dimensional grounds, we expect that
\begin{eqnarray}
 {{dN_{C\to DY}} \over {d^4x dydM^2}} = \alpha^2 {{(\alpha_s n_{gluon})^3} \over M_{Debye}^7},
 g^{\prime\prime}(M/M_{Debye})
\end{eqnarray}
where we have assumed that the condensate density is of the order of the gluon number density as in Eq.(\ref{eqn_ng_nc})and that the typical scale
for the energy of gluons in the condensate is of order the Debye mass.  Integration over time leads to
  \begin{equation} \label{eqn_dilep_int}
  {{dN_{C\to DY}} \over {dydM^2}} \sim \alpha^2 R_0^{\prime 2} N_{part}^{2/3} \left( {Q_{sat} \over M} \right)^{\eta^{\prime}}
\end{equation}
where
 \begin{equation}
        \eta^\prime_{perturbative} = \frac{9(3-\delta)}{5+3\delta}
 \end{equation}

\begin{figure}[h!]
\vspace{0.0cm}
\begin{center}
\scalebox{1.0}[1.0] { \hspace{0.2cm}
  \includegraphics[scale=.80]{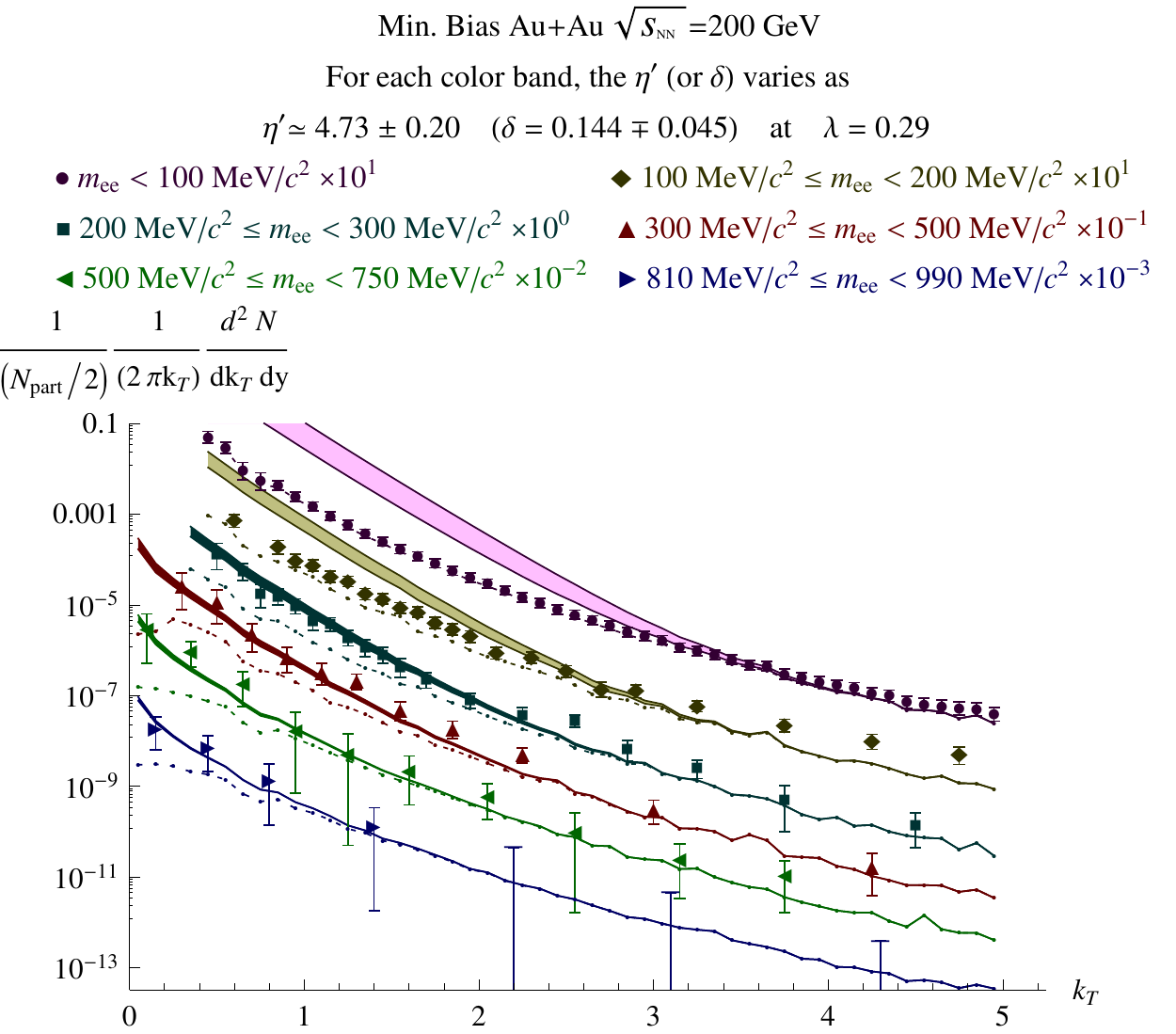} }
\end{center}
\caption{(Color online) The comparison between the PHENIX dilepton data and the model \cite{Chiu2012}.} \label{dileptons} \vspace{0.2cm}
\end{figure}

The analytical results are compared to RHIC data in Fig.~\ref{dileptons}.

From Fig.~\ref{dileptons} we see that our simple model provides a good description of the experimental data.

\section{Turbulent plasma}

In this section we consider the key physical characteristics of turbulent ultrarelativistic plasma, its polarization properties in the Abelian
\cite{KLM2012a} and non-Abelian \cite{KLM2012b} cases as well as its anomalous viscosity \cite{ABM06,ABM07,ABM11} and jet quenching
\cite{MMW2007}.

\subsection{Turbulent polarization: QED plasma}

Let us first consider polarization properties of the turbulent ultrarelativistic QED plasma \cite{KLM2012a}.  A weakly turbulent plasma is
described as perturbation of an equilibrated system of (quasi-)particles by weak turbulent fields $F^T_{\mu \nu}$. In the collisionless Vlasov
approximation we employ , the plasma properties are defined by the following system of equations ($F^R_{\mu \nu}$ is a regular non-turbulent
field):
\begin{eqnarray}
&& p^{\mu}\left[ \partial_{\mu}- e q \left( F^R_{\mu \nu} + F^T_{\mu \nu} \right) \dfrac{\partial}{\partial p_{\nu}}\right]f(p,x,q)=0 \nonumber\\
&&\partial^{\mu}\left( F^R_{\mu \nu} + F^T_{\mu \nu} \right)= j_{\nu}(x) = e \sum_{q,s}\int dp\, p_{\nu}\, q\, f(p,x,q) . \label{kinetic+maxw}
\end{eqnarray}
The stochastic ensemble of turbulent fields is assumed to be Gaussian and characterized by the following correlators:
\begin{equation}
\langle F_{\mu \nu}^{T}\rangle=0, \;\;\;\;\;  \langle F^{T \mu \nu}(x)F^{T \mu^{\prime} \nu^{\prime}}(y)\rangle=K^{\mu \nu \mu^{\prime} .
\nu^{\prime}}(x,y) \label{turbens}
\end{equation}
Following \cite{ABM} we use the following parametrization of $K^{\mu \nu \mu^{\prime} \nu^{\prime}}(x,y) $ :
\begin{equation}
K^{\mu \nu \mu^{\prime} \nu^{\prime}}(x)=K_{0}^{\mu \nu \mu^{\prime} \nu^{\prime}}\exp\left[-\dfrac{t^{2}}{2\tau^{2}}-\dfrac{r^{2}}{2 a^{2}}
\right]
\end{equation}

By definition, turbulent polarization is defined as a response to a regular perturbation that depends on turbulent fields. In the linear response
approximation it is fully described by the polarization tensor $\Pi^{\mu \nu} (k)$ which can be computed by taking a variational derivative of the
averaged induced current $\langle j^{\mu}( k \; \vert F^R,F^T) \rangle_{F^T}$ over the regular gauge potential $A^R_{\nu}$:
\begin{eqnarray}
&&\Pi^{\mu \nu} (k) = \frac{ \delta \langle   j^{\mu}(k \vert F^R,F^T) \rangle_{F^T}}{\delta A^R_{\nu}} \label{PolarTensDef}  \\
&&\langle   j^{\mu}( k \; \vert F^R,F^T) \rangle_{F^T} = e \sum_{q,s} \int d P p_{\nu} q \langle \delta f (p,k,q \vert F^R,F^T) \rangle_{F^T}
\label{incur}
\end{eqnarray}

To organize the calculation in the efficient way it is useful to rewrite (\ref{kinetic+maxw}) as follows:
\begin{equation}
f=f^{eq} + G p^{\mu}F_{\mu\nu}\partial_{p}^{\mu}f \; , \;\;\;\; G \equiv \dfrac{e q}{\imath((pk)+\imath \epsilon)},
\end{equation}
where $f^{eq}$ is a distribution function characterizing the original non-turbulent plasma and introduce the following systematic expansion in the
turbulent and regular fields:
\begin{equation}
\delta f = \sum_{m=0}\sum_{n=0}\rho^{m}\tau^{n}\delta f_{mn}, \;\;\;  F^{\mu \nu} =  \sum_{m=0}\sum_{n=0}\rho^{m}\tau^{n}F_{mn}^{\mu\nu},
\end{equation}
where powers of $\rho$ count those of $F^R$ and powers of $\tau$ count those of $F^T$. Turbulent polarization is described by contributions of the
first order in the regular and the second in the turbulent fields. The lowest nontrivial contribution to the induced current (\ref{incur}) is thus
given by $\delta f_{12}$:
\begin{equation}
\delta f  \simeq  \delta f_{\rm HTL} + \langle \delta f_{12} \rangle_{\rm I}  + \langle \delta f_{12} \rangle_{\rm II} \;  \nonumber
\end{equation}
where
\begin{eqnarray}
\delta f_{\rm HTL} & = & Gp_{\mu}F_{10}^{\mu\nu}\partial_{\mu,p}f^{\rm eq} \label{fhtl}  \nonumber \\
 \langle \delta f_{12} \rangle_{\rm I} & = & Gp_{\mu} \langle F_{01}^{\mu\nu}\partial_{\nu,p}Gp_{\mu^{\prime}}
 F_{10}^{\mu^{\prime}\nu^{\prime}}\partial_{\nu^{\prime},p}Gp_{\rho} F_{01}^{\rho
 \sigma}\rangle \partial_{\sigma,p} f^{\rm eq} \label{f12I} \nonumber \\
\langle \delta f_{12} \rangle_{\rm II} & = & Gp_{\mu} \langle  F_{01}^{\mu \nu}\partial_{\nu,p}Gp_{\mu^{\prime}}
F_{01}^{\mu^{\prime}\nu^{\prime}}\partial_{\nu^{\prime},p}Gp_{\rho} F_{10}^{\rho \sigma}
 \rangle \partial_{\sigma,p} f^{\rm eq} \label{f12II} \nonumber
\end{eqnarray}

Generically one has the following decomposition of the polarization tensor:
\begin{equation}
\Pi_{i j}(\omega,\mathbf k \, ; l)=\left(\delta_{i j}-\dfrac{k_{i}k_{j}}{k^{2}}\right)\Pi_{T}(\omega,\absvec{k}\, ; l)
+\dfrac{k_{i}k_{j}}{k^{2}}\Pi_{L} (\omega,\absvec{k} \, ; l) \label{poltendec}
\end{equation}
where $l \equiv \sqrt{2} (\tau a) / \sqrt{\tau^2+a^2}$. Its components can be rewritten as sums of theh leading Hard Thermal Loops (HTL)
contributions and the gradient expansion in the scale of turbulent fluctuations $l$:
\begin{eqnarray}
&& \Pi_{L(T)} (\omega,\mathbf{k} \, ; l)  =   \Pi^{\; \rm HTL}_{L(T)} (\omega,\mathbf{k}) +  \Pi^{\; \rm turb}_{L(T)} (\omega,\mathbf{k} \vert \;
l) \label{poltenregturb} \\
&& \Pi^{\; \rm turb}_{L(T)} (\omega,\absvec{k} \, ; l)  =  \sum_{n=1}^\infty \dfrac{(\absvec{k}l)^n}{\mathbf{k}^2}
 \left [ \phi^{\; (n)}_ {L(T)} (x) \langle E_{\rm turb}^2 \rangle +
\chi^{\; (n)}_{L(T)} (x) \langle B_{\rm turb}^2 \rangle \right] \label{gradexp} \nonumber
\end{eqnarray}
where $x=\omega/\absvec{k}$ and the HTL contributions to the polarization tensor read
\begin{eqnarray}
&&\Pi_{L }^{\mathrm HTL} (\omega,\absvec{k}) =- m^2_D x^2 \left[1-\dfrac{x}{2} \; L(x) \right], \nonumber \\
&&\Pi_{T}^{\mathrm HTL} (\omega,\absvec{k})= m^2_D \dfrac{x^2}{2} \left[1+\dfrac{1}{2 x} \; (1-x^2 )\; L(x) \right] \nonumber \\
&& L(x) \equiv \ln\left|\dfrac{1+x}{1-x}\right|-\imath\pi\theta(1-x); \;\;\; m^2_D=e^2 T^2/3 . \label{HTL}
\end{eqnarray}
The computation of turbulent polarization was carried out to second order in the gradient expansion \cite{KLM2012b}. To the leading order in the
gradient expansion one gets
\begin{eqnarray}
 \phi_{\rm I\;T}^{\; (1)} (x) & = &\frac{\imath e^{4}}{6\pi\sqrt{\pi}} \; 2 x \left[\dfrac{4+10 x^{2}-6 x^{4}}{3(1-x^{2})}+x(1-x^{2}) \; L(x) \right] \label{phi1T} \\
 \phi_{\rm I\;L}^{\; (1)} (x) & = & -\frac{\imath e^{4}}{6\pi\sqrt{\pi}}\; \dfrac{8 x^{3}}{3(1-x^{2})^{2}} \label{phi1L}
\end{eqnarray}
and
\begin{eqnarray}
 \chi_{\rm I\;T}^{\; (1)} (x) & = & \frac{\imath e^{4}}{6\pi\sqrt{\pi}}\; 4 x \left[ \dfrac{-2+6 x^{2}}{3(1-x^{2})}+x \; L(x)
 \right]\label{chi1T} \\
 \chi_{\rm I\;L}^{\; (1)} (x) & = & - \frac{\imath e^{4}}{6\pi\sqrt{\pi}} \; \dfrac{8 x^{3}}{3(1-x^{2})^{2}} \label{chi1L}.
\end{eqnarray}

Let us first discuss the turbulent contributions to the imaginary part of the polarization tensor \cite{KLM2012b}. These can be summarized as
follows. The sign of the imaginary part of the turbulent contribution to the polarization operator in the timelike domain $x>1$ is negative and
corresponds to turbulent damping of timelike collective excitations. This refers to both transverse and longitudinal modes. As the HTL
contribution in this domain is absent, this turbulent damping is a universal phenomenon present for all $\omega,k$ such that $\omega > k$ and all
values of the parameters involved ($l$, $\langle B^2 \rangle$, $\langle E^2 \rangle$). The turbulent damping leads to an attenuation of the
propagation of collective excitations at some characteristic distance. The situation in the spacelike domain $x<1$ is more diverse. In contrast
with the timelike domain the gradient expansion for the imaginary part of the polarization tensor starts from the negative HTL contribution
corresponding to Landau damping. The imaginary parts of turbulent contributions to the longitudinal polarization tensor are negative and are thus
amplifying the Landau damping. The most interesting contributions come from the turbulent contributions to the transverse polarization tensor. The
electric contribution ${\mathrm{Im}}[\phi_T^{\; (1)} (x)]$ in the spacelike domain is positive at all $x$ while the magnetic contribution
${\mathrm{Im}}[\chi_T^{\; (1)} (x)]$ is negative for $x<x^* \approx 0.43$ and positive for $x>x^*$. This means that the turbulent plasma becomes
unstable for sufficiently strong turbulent fields.

It is also of interest to analyze the effects of turbulence on the properties of collective excitations of QED plasma, the plasmons
\cite{KLM2012b}. The plasmons are characterized by dispersion relations $\omega_{\rm T(L)} (\absvec{k})$ that are read from the solutions of
dispersion equation for the corresponding components of dielectric permittivity, which are just a real part  of zeroes of inverse transverse and
longitudinal wave propagators:
\begin{equation}
\begin{split}
&Re\left[ \left. {\mathbf k}^{2}\left(1-\dfrac{\Pi_{\rm L}(k^{0},\absvec{k})}{\omega^{2}}\right)\right|_{k^{0}=\omega_{\rm L}(\absvec{k})}\right]=0\\
&Re\left[{\mathbf k}^{2}-(k^{0})^{2}+\Pi_{\rm T}((k^{0},\absvec{k})\mid_{k^{0}=\omega_{\rm T}(\absvec{k})}\right]=0
\end{split}
\end{equation}
Thus, real part of polarization tensor corresponds to propagation of plasmons in a medium, while it's imaginary part defies plasmon smearing.

Let us focus first on a shift of plasmons dispersion relations in turbulent medium. In general dispersion equations can be solved only
numerically. Analytical expressions can be obtained in certain limits. Let us focus on the deeply timelike regime of $x \gg 1$.  In non-turbulent
HTL Vlasov plasma the time-like plasmon modes do not decay, since imaginary part of polarization tensor in that limit is zero. For frequencies
$\frac{k}{\omega_{pl}}<<1$ the corresponding solutions of dispersion equations may be expanded as powers of $\frac{\absvec{k}}{\omega_{\rm pl}}$:
\begin{equation}
\begin{split}
 & \omega_{\rm L}^{2}(\absvec{k})_{\rm HTL}=\omega_{\rm pl}^{2}\left(1+\dfrac{3}{5}\left(\dfrac{\absvec{k}}{\omega_{\rm pl}}\right)^{2}+
 O\left(\left(\dfrac{\absvec{k}}{\omega_{\rm pl}}\right)^{4}\right)\right)\\
& \omega_{\rm T}^{2}(\absvec{k})_{\rm HTL} =\omega_{\rm pl}^{2} \left(1+\dfrac{6}{5}\left(\dfrac{\absvec{k}}{\omega_{\rm
pl}}\right)^{2}+O\left(\left(\dfrac{\absvec{k}}{\omega_{\rm pl}}\right)^{4}\right)\right)
\end{split}
\label{disprelHTL}
\end{equation}
where we have used a standard definition for the plasma frequency $\omega^2_{\rm pl}=m^2_D/3$.

In a turbulent plasma plasmons decay even in a Vlasov limit since polarization tensor has imaginary part. As to the turbulent modifications of the
HTL dispersion relation (\ref{disprelHTL}), it can be conveniently written as
\begin{equation}
\begin{split}
 &\omega_{\rm L}^2(\absvec{k})_{\rm turb}=(\omega^{\rm turb}_{\rm pl\; L})^2
 \left(1+\dfrac{3}{5}y_{\rm L}^2\right)- \dfrac{e^{4} l^{2}}{6 \pi^{2}}\left(\dfrac{24}{5}\langle E^{2}\rangle+\dfrac{64}{15}\langle B^{2}\rangle \right)y_{\rm L}^2+O\left(y_{\rm L}^4 \right)\\
 &\omega_{\rm T}^2(\absvec{k})_{\rm turb}=(\omega^{\rm turb}_{\rm pl\;T})^2\left(1+\dfrac{3}{5}y_{\rm T}^2\right)-\dfrac{e^{4} l^{2}}{6 \pi^{2}} \left( \dfrac{24}{7}\langle
E^{2}\rangle+\dfrac{32}{15}\langle B^{2}\rangle\ \right) y_{\rm T}^2+O\left(y_{\rm T}^4 \right)\; ,
\end{split}
\end{equation}
where
\begin{equation}
y_{\rm L}=\frac{\absvec{k}}{\omega^{\rm turb}_{\rm pl\;L}}; \;\;\;\; y_{\rm T}=\frac{\absvec{k}}{\omega^{\rm turb}_{\rm pl\;T}}\;,
\end{equation}
and
\begin{equation}
\begin{split}
&(\omega^{\rm turb}_{\rm pl\;L})^{2}=\omega_{\rm pl\;L }^{2}-\dfrac{e^{4} l^{2}}{6 \pi^{2}} \left(\dfrac{16}{3}\langle E^{2}\rangle+\dfrac{8}{3}\langle B^{2}\rangle\right)\\
&(\omega^{\rm turb}_{\rm pl\;T})^{2}=\omega_{\rm pl\;T }^{2}-\dfrac{e^{4} l^{2}}{6 \pi^{2}} \left(\dfrac{128}{15}\langle
E^{2}\rangle+\dfrac{8}{3}\langle B^{2}\rangle\right).
\end{split}
\end{equation}

Now let us consider plasmons smearing. As it can be easily seen that a rate of decay for plasmons is connected to an imaginary part of
polarization tensor by a formula:
\begin{equation}
\Gamma_{T(L)}=\sqrt{-Im(\Pi_{T(L)})}
\end{equation}
In a timeline region considered above imaginary part of both transverse and longitudinal components of polarization tensor are lesser than zero:
there i no instability for timeline modes. Also it should be noted that turbulent smearing is a leading order effect on $(k l)$ compared with a
turbulent modification plasmon dispersion relations.

\subsection{Turbulent polarization: QCD plasma}

Let us now discuss a generalization of the results of \cite{KLM2012a,KLM2012b} on the polarization properties on the non-Abelian QCD plasma.

The generalization of Eq.~(\ref{kinetic+maxw}) to the non-Abelian case reads
\begin{equation}
p^{\mu} \left[\partial_{\mu}-g f_{a b c}A_{\mu}^{b} Q^{c} \dfrac{\partial}{\partial_{Q^{a}}} - g Q_{a} F_{\mu
\nu}^{a}\dfrac{\partial}{\partial_{p_{\nu}}}\right]=0, \label{kinetic}
\end{equation}
where the fields $F_{\mu \nu}$ satisfy the Yangh-Mills equations
\begin{equation}
D^{\mu}F_{\mu \nu}^{ a}=j_{\nu}^{a} \label{MaxwYM}
\end{equation}
The main distinction from the Abelian case is the dependence of the distribution function on the color spin $Q$, where for $SU(3)$
$Q=(Q^{1},Q^{2},...,Q^{8})$, so that $f(x,p,Q)$. The components of color spin $Q=(Q^{1},Q^{2},...,Q^{8})$ are dynamic variables satisfying the
Wong equation
\begin{equation}
\dfrac{d Q^{a}}{d \tau}=-g f^{a b c}p^{\mu}A_{\mu}^{b}Q^{c} \label{WongEq}
\end{equation}
which, with Eqs.~(\ref{kinetic},\ref{MaxwYM}), completes the dynamical description of QCD plasma.

The description of the properties of turbulent QCD plasma is based on separating the regular and turbulent contributions ot the distribution
functions and gauge potentials $A_{a}^{\mu}$:
\begin{equation}
f=f^{R}+f^{T}, \;\;\;\; A_{\mu}^{a}=A_{\mu}^{R a}+A_{\mu}^{T a},
\end{equation}
where we assume that $\left\langle A_{\mu}^{a}\right\rangle=A_{\mu}^{R a}$ and $\left\langle A_{\mu}^{T a}\right\rangle=0$

It is possible to define gauge transformations of regular and turbulent gauge potentials in such a way that
\begin{equation}
\begin{split}
&\delta A_{\mu}^{R a}=\partial_{\mu}\alpha^{a}+g f^{a b c}A_{\mu}^{R b}\alpha^{c}\\
&\delta A_{\mu}^{T b}=g f^{a b c}A_{\mu}^{T b} \alpha^{c} \label{bfgauge}
\end{split}
\end{equation}
which is technically equivalent to choosing the background field gauge. A very useful property following from this choice is that the basic
property of turbulent fields  $\left\langle A_{\mu}^{T a}\right\rangle=0$ is gauge invariant. The corresponding decomposition of gauge field
strength reads
\begin{equation}
F_{\mu \nu}^{a}=F_{\mu \nu}^{R a}+\mathbf{F}_{\mu \nu}^{T a}+ \EuScript{F}_{\mu \nu}^{T a}
\end{equation}
where:
\begin{equation}
\begin{split}
&F_{\mu \nu}^{R a}= \partial_{\mu}A_{\nu}^{R a}-\partial_{\nu}A_{\mu}^{R a}+g f^{abc}A_{\mu}^{R b}A_{\nu}^{R c}\\
&\EuScript{F}_{\mu\nu}^{T a}= \partial_{\mu}A_{\nu}^{T a}-\partial_{\nu}A_{\mu}^{T a}+g f^{abc}A_{\mu}^{T b}A_{\nu}^{T c}\\
&\mathbf{F}_{\mu \nu}^{T a}=g f^{a b c}\left( A_{\mu}{T b}A_{\nu}^{R c}+A_{\mu}{R b}A_{\nu}^{T c}\right)
\end{split}
\end{equation}

The computation of polarization properties proceeds through expanding $f^{R}$,  $f^{T}$ in a series in the regular potential ($f^{(0)}\sim
(A{R})^{0}$, $f^{(1)}\sim (A{R})^{1}$, ...). Assuming $f^{R (0)}(x,p,Q)=f^{eq}(p)$ we have
\begin{equation}
(p^{\mu}\partial_{\mu})f^{T (0)}=g  p^{\mu} Q_{a}\EuScript{F}_{\mu\nu}^{T a} \dfrac{\partial}{\partial p^{\nu}}f^{R (0)} \label{ft0}
\end{equation}
The corresponding equations for the first order turbulent contributions to the distribution function read
\begin{equation}
\begin{split}
&(p^{\mu} \partial_{\mu})f^{T (1)}= g p^{\mu} f^{a b c} A_{\mu}^{T b}Q^{c}\dfrac{\partial}{\partial Q^{a}}f^{R (1)}+g^{2}p^{\mu}f^{a b c}
A_{\mu}^{R b} Q^{c}\dfrac{1}{(p \partial)}p^{\mu^{\prime}}\EuScript{F}_{\mu^{\prime} \nu}^{T a}\dfrac{\partial}{\partial p_{\nu}}f^{R (0)}
+\\
&g p^{\mu} Q_{a} \mathbf{F}_{\mu\nu}^{T a}\dfrac{\partial}{\partial_{\nu}}f^{R (0)}+g p^{\mu} Q_{a}\EuScript{F}_{\mu \nu}^{T a}
\dfrac{\partial}{\partial p_{\nu}}f^{R (1)}+g^{2}p^{\mu}p^{\mu}p^{\prime}Q^{a}F_{\mu\nu}^{R a}\dfrac{\partial}{\partial
p_{\nu}}\dfrac{1}{(p^{\mu}\partial_{\mu})}Q_{c}\EuScript{F}_{\mu^{\prime}\nu^{\prime}}^{T c}\dfrac{\partial}{\partial p_{\nu^{\prime}}} f^{R (0)}+
\\ & g^{2}p^{\mu}f^{a b c}A_{\mu}^{R b}Q^{c}\dfrac{\partial}{\partial
Q^{a}}\dfrac{1}{(p^{\mu}\partial_{\mu})}p^{\mu^{\prime}}_{d}\EuScript{F}_{\mu^{\prime}\nu}^{T d}\dfrac{\partial}{\partial p_{nu}}f^{R (0)}
\end{split}
\label{ft1}
\end{equation}
and
\begin{equation}
\begin{split}
&(p^{\mu}\partial_{\mu})f^{R (1)}= g p^{\mu}f^{a b c}\left\langle A_{\mu}^{T b}Q^{c}\dfrac{\partial}{Q^{a}}f^{T (1)}\right\rangle+ g p^{\mu} Q_{a}\left\langle\EuScript{F}_{\mu\nu}^{T a}\dfrac{\partial}{\partial p_{\nu}}f^{T (1)}\right\rangle+g p^{\mu}Q_{a}\left\langle\mathbf{F}_{\mu \nu}^{R}\dfrac{\partial}{\partial p_{\nu}}f^{T(0)}\right\rangle\\
&+g p^{\mu}Q_{a}F_{\mu \nu}^{R a}\dfrac{\partial}{\partial p_{\nu}}f^{R (0)}
\end{split}
\label{fr1}
\end{equation}

Substituting (\ref{ft0}) and (\ref{ft1}) to (\ref{fr1}) one arrives at the final expression for the first order regular correction to the
distribution function. A detailed analysis shows that in the relevant long wavelength limit we are left with only two contributions:
\begin{equation}
f^{R (1)}=HTL + I_{1}+I_{2}
\end{equation}
where
\begin{equation}
\begin{split}
&I_{1}=g^{3}p^{\mu}  f^{a b c} \left\langle A_{\mu}^{T b} Q^{c} \dfrac{\partial}{\partial Q^{a}} p^{\mu^{\prime}} \dfrac{1}{p^{\mu} \partial_{\mu}}
f^{d e f}A_{\mu^{\prime}}^{R e}Q^{f}\dfrac{\partial}{\partial Q^{d}}\dfrac{1}{p^{\mu}\partial_{\mu}}p^{\mu^{\prime}}
Q_{g}\EuScript{F}_{\mu^{\prime \prime} \nu}\right\rangle\dfrac{\partial}{\partial p_{\nu}}f^{R(0)}\\
&I_{2}=g^{2}p^{\mu}f^{a b c} \left\langle A_{\mu}^{T b} Q^{c}\dfrac{\partial}{\partial Q^{a}}\dfrac{1}{(p^{\mu} \partial_{\mu})}
p^{\mu^{\prime}}Q_{d}\mathbf{F}_{\mu^{\prime} \nu}^{T d}\right\rangle\dfrac{\partial}{\partial p_{\nu}}f^{R (0)}\\
\end{split}
\end{equation}
Averaging over the stochastic color fields involves two two-point correlators
\begin{equation}
\left\langle A_{\mu}^{T a}(x) A_{\nu}^{T b}(y) \right\rangle = G_{\mu \nu}^{a b}(x,y)
\end{equation}
and
\begin{equation}
\left\langle \EuScript{F}_{\mu\nu}^{T a}(x) U^{a b}(x,y) \EuScript{F}_{\mu^{\prime} \nu^{\prime}}^{T b}(y)\right\rangle= K_{\mu \nu \mu^{\prime}
\nu^{\prime}}^{a b}(x,y)
\end{equation}
We shall restrict our consideration to the plasma which is on average homogeneous, $K_{\mu \nu \mu^{\prime} \nu^{\prime}}^{T a} (x,y)=K_{\mu \nu
\mu^{\prime} \nu^{\prime}}^{T a} (x-y)$ (same for $G_{\mu \nu}^{a b}$), and assume that the stichastic correlators are symmetric under
permutations of both color and Lorentz indices.

Let us chose following explicit parametrization for the non-Abelian correlation function $G_{\mu \nu}^{a b}$:
\begin{equation}
G_{\mu \nu}^{a b} = \delta_{a b}\left[ g_{\mu \nu} g_{\nu 0}\left\langle A_{0}^{2} \right\rangle+\dfrac{1}{3}\hat{\delta}_{\mu \nu} \left\langle
\mathbf{A}^{2} \right\rangle \right]\exp\left[ -\dfrac{r^{2}}{2 a^{2}}-\dfrac{t^2}{2 \tau^{2}}\right]
\end{equation}
Defining $f^{R a (1)}(x,p)=\int Q^{a}\, d\, Q\; f^{R (1)} (x,p,Q)$ we get
\begin{equation}
[(p^{\mu}\partial_{\mu}) +p \gamma]f^{R l (1)}=\int Q^{l}\, d \,Q \; \left(HTL+I_{1} +I_{2} \right), \label{RenProp}
\end{equation}
where
\begin{equation}
\gamma=g^{2} \dfrac{N^{2}-1}{4 N}\sqrt{\pi} l \left[\left\langle A_{0}^{2}\right\rangle + \left\langle\dfrac{1}{3} \mathbf{A}^{2}
\right\rangle\right]
\end{equation}
and $ l=\frac{1}{\sqrt{\frac{1}{2 a^{2}}+\frac{1}{2 \tau^{2}}}}$. Let us stress that only the sum of contributions from $HTL$, $I_{1}$ and $I_{2}$
is gauge invariant.

Detailed calculations show that to the leading order there is no contribution from the non-Abelian correlator $G_{\mu \nu}^{a b}$, so that the
only modification distinguishing QCD plasma from the QED one is an overall normalization so that
\begin{eqnarray}
\phi_{T,L}(x) & \to & C_{q(g)} \phi_{T,L}(x) \nonumber \\
\chi_{T,L}(x) & \to & C_{q(g)} \chi_{T,L}(x),
\end{eqnarray}
where for quarks
\begin{equation}
C_{q}=g^{4} N_{q}\dfrac{N^{2}-1}{4 N}
\end{equation}
and for gluons
\begin{equation}
C_{g} = \dfrac{2 g^{3} N^{2}}{N+\frac{N_{q}}{2}}
\end{equation}
where we have taken into account a necessity of introducing an infrared cutoff at $m_D$ when computing the integral $\int dp \dfrac{df}{dp}$ for
massless bosons.

\subsection{Turbulent anomalous viscosity}

One of the most important requirements for any scenario describing the relevant physics of the early stage of nuclear collisions is its ability of
explaining an (effectively) small viscosity characterizing collective expansion of dense matter created in these collisions. Let us recall the
standard kinetic theory expression for shear viscosity
\begin{equation}
\eta = \frac{1}{3} n\, \langle p \rangle_T\, \lambda_f ,
\end{equation}
where $n$ is the density of the medium, $\langle p \rangle_T$ is the thermal momentum of particles and $\lambda_f$ is the mean free path. In the
standard perturbative case the source of $\lambda_f$ is perturbative scattering which is parametrically weak and, therefore, leads to very large
values of $\lambda_f$ and, consequently, shear viscosity $\eta$. The situation is dramatically different in turbulent plasmas
\cite{ABM06,ABM07,ABM11}, where particle scatters on strong turbulent fields. The physical picture is here that of plasma particles scattering on
domains of coherent turbulent fields, the size of the domains being controlled by the corresponding correlation length $l$, so that
\begin{equation}
\lambda_f = \frac{\langle p \rangle_T^2}{g^2\, Q^2 \langle B^2 \rangle\, l} .
\end{equation}
leading to the anomalously small turbulent shear viscosity \cite{ABM06,ABM07,ABM11}
\begin{equation}
\eta_{\rm A} \approx \frac{\frac{9}{4} s\, T^3}{g^2\, Q^2 \langle B^2 \rangle\, l}
\end{equation}
valid for the nearly equilibrated plasma.

\subsection{Turbulent jet quenching}

Another very attractive feature of the turbulent plasma scenario is its natural ability to describe the observed strong jet quenching, i.e. large
energy loss experienced by fast particles propagating through early QCD matter \cite{MMW2007}. Indeed, it is easy to calculate transverse
broadening $\langle (\Delta p_\perp)^2 \rangle$ induced by the same process of particle scattering on turbulent field domains and the
corresponding (anomalous) jet quenching parameter $\hat{q}_A$:
\begin{equation}
\hat{q}_{\rm A} = g^2\, Q^2 \langle B^2 \rangle\, l.
\end{equation}
leading, in particular, to an interesting relation between the two above-discussed anomalous quantities \cite{MMW2007}:
\begin{equation}
\label{eq:eta-qhat} \frac{\eta_{\rm A}}{s} \propto \frac{T^3}{\hat{q}_{\rm A}} ,
\end{equation}
i.e. the smaller is the anomalous turbulent viscosity, the large is the anomalous turbulent jet quenching.

\section{Conclusions}

With a wealth of new experimental data and theoretical ideas the physics of high energy heavy ion collisions remains a truly exciting domain of
research. We are sure that in the near future we shall see rapid progress both in the quality of expreimental data and that od theoretical
resuarch.


\begin{thebibliography}{99}


\bibitem{DL10} I.M. Dremin, A.V. Leonidov, {\it Physics-Uspekhi} {\bf 53} (2010), 1123


\bibitem{hydroRHIC}
I. Arsene, et al. (BRAHMS), {\it Nucl. Phys.} {\bf A757} (2005), 1 \\
B.B. Back, et al. (PHOBOS), {\it Nucl. Phys.} {\bf A757} (2005), 28 \\
J. Adams, et al. (STAR), {\it Nucl. Phys.} {\bf A757} (2005), 102 \\
K. Adcox, et al. (PHENIX), {\it Nucl. Phys.} {\bf A757} (2005), 184


\bibitem{hydroLHC}
K. Aamodt, et al. (ALICE), {\it Phys. Rev. Lett.} {\bf 105} (2010), 252302 \\
G. Aad, et al. (ATLAS), {\it Phys. Lett.} {\bf B707} (2012), 330 \\
CMS Collaboration, CMS PAS HIN-10-002


\bibitem{H05}
U.W. Heinz, Thermalization at RHIC, {\it AIP Conf. Proc.} {\bf 739} (2005), 163-180, arXiv:nucl-th/0407067


\bibitem{MSW12}
B. Muller, J. Schukraft, B. Wyslouch, {\it Annu. Rev. Nucl. Part. Sci.} {\bf 62} (2012), 361, arXiv:1202.3233


\bibitem{GJSTV13}
C. Gale, S. Jeon, B. Schenke, P. Tribedy, R. Venugopalan, {\it Phys. Rev. Lett.} {\bf 110} (2013), 012302

\bibitem{ahydro}
R. Ryblewski, \emph{Collective phenomena in the early stages of relativistic heavy ion collisions}, arXiv:1297.0629 \\
W. Florkowski, M. Martinez, R. Ryblewski, M. Strickland, \emph{Anisotropic hydrodynamics - basic concepts}, arXiv:1301.7593

\bibitem{janik}
M. Heller, R.A. Janik, P. Witaszyk, {\it Phys. Rev. Lett.} {\bf 108} (2012), 201602 \\
M. Heller, R.A. Janik, P. Witaszyk, {\it Phys. Rev.} {\bf D85} (2012), 126002



\bibitem{KMW95}
A. Kovner, L.D. McLerran and H. Weigert, {\it Phys. Rev.} {\bf D52} (1995), 6231-6237
\bibitem{LM06}
T. Lappi, L. McLerran, {\it Nucl. Phys.} {\bf A772} (2006), 200-212


\bibitem{Blaizot2012}
J.-P. Blaizot, F. Gelis, J. Liao, L. McLerran, R. Venugopalan, {\it Nucl. Phys.} {\bf A873} (2012), 68

\bibitem{Chiu2012}
M. Chiu, T.K. Hemmick, V. Khachatryan, A. Leonidov, J. Liao, L. McLerran, \textit{Production of Photons and Dileptons in the Glasma},
ArXiv:1202.3679


\bibitem{RV06}
P. Romatschke, R. Venugopalan, {\it Phys. Rev. Lett.}\; {\bf 96} (2006), 062302; {\it Eur. Phys. J.}\; {\bf A29} (2006), 71; {\it Phys. Rev.}\;
{\bf D74} (2006), 045011
\bibitem{FG12}
K. Fukushima, F. Gelis, {\it Nucl. Phys.}\; {\bf A874} (2012), 108


\bibitem{DEGV11}
K. Dusling, T. Epelbaum, F. Gelis and R. Venugopalan, {\it Nucl. Phys.} {\bf A850} (2011), 69-109
\bibitem{EG11}
T. Epelbaum, F. Gelis, {\it Nucl. Phys.} {\bf A872} (2011), 210-244
\bibitem{DEGV12}
K. Dusling, T. Epelbaum, F. Gelis and R. Venugopalan, {\it Phys. Rev.} {\bf D86} (2012), 085040


\bibitem{W59}
E.S. Weibel, {\it Phys. Rev. Lett.}\; {\bf 2:83} (1959)
\bibitem{M88}
S. Mrowczynski, {\it Phys. Lett.}\;  {\bf B214} (1988), 587.
\bibitem{PS88}
Y.E. Pokrovsky, A.V. Selikhov, {\it JETP Lett.}\; {\bf 47} (1988), 12-14
\bibitem{M93}
S. Mrowczynski, {\it Phys. Lett.}\;  {\bf B314:118} (1993)
\bibitem{M97}
S. Mrowczynski, {\it Phys. Lett.}\;  {\bf B393:26} (1997).
\bibitem{ALM03}
P. Arnold, J. Lenaghan, G.D. Moore, {\it JHEP} {\bf 08} (2003), 002
\bibitem{ALMY05}
P. Arnold, J. Lenaghan, G.D. Moore, L.G. Yaffe, {\it Phys. Rev. Lett.}\; {\bf 94} (2005), 072302



\bibitem{AM06a}
P. Arnold, G.D. Moore, {\it Phys. Rev.} {\bf D73} (2006), 025006
\bibitem{AM06b}
P. Arnold, G.D. Moore, {\it Phys. Rev.} {\bf D73} (2006), 025013


\bibitem{R09}
A. Rebhan, \textit{Hard loop effective theory of the (anisotropic) quark gluon plasma}, arXiv:0811.0457 [hep-ph]
\bibitem{KM11}
A. Kurkela, G.D. Moore, {\it JHEP} {\bf 1112} (2011), 044
\bibitem{KM12}
A. Kurkela, G.D. Moore, {\it JHEP} {\bf 1204} (2012), 120
\bibitem{IRS11}
A. Ipp, A. Rebhan and M. Strickland, {\it Phys. Rev.} {\bf D84} (2011), 056003
\bibitem{CR11}
M.E. Carrington, A. Rebhan, \textit{Perturbative and Nonperturbative Kolmogorov Turbulence in a Gluon Plasma}, arXiv:1012.0298 [hep-ph]


\bibitem{BSS07}
J. Berges, S. Scheffler and D. Sexty, {\it Phys. Rev.} {\bf D77} (2008), 034504
\bibitem{BGSS09}
J. Berges, D. Gelfand, S. Scheffer and D. Sexty, {\it Phys. Lett.} {\bf B677} (2009), 210
\bibitem{BSS09}
J. Berges, S. Scheffler and D. Sexty, {\it Phys. Lett.} {\bf B681} (2009), 362
\bibitem{MSW07}
A.H. Mueller, A.I. Soshi and S.M.H. Wong, {it Nucl. Phys.} {\bf B760} (2007), 145-165


\bibitem{KMOSTY10}
T. Kunihiro, B. M\"{u}ller, A. Ohnishi, A. Sch\"{a}fer, T. Takahashi, A. Yamamoto, {\it Phys. Rev.}\; {\bf D82} (2010), 114015


\bibitem{ABM06}
M. Asakawa, S. A. Bass, B. M\"{u}ller, {\it Phys. Rev. Lett.}\; {\bf 96} (2006), 252301
\bibitem{ABM07}
M. Asakawa, S. A. Bass, B. M\"{u}ller, {\it Progr. Theor. Phys.}\; {\bf 116} (2007), 725
\bibitem{ABM11}
M. Asakawa, S. A. Bass, B. M\"{u}ller, {\it Nucl.Phys.}\; {\bf A854} (2011), 76


\bibitem{Tsit}
V. N. Tsytovich \textit{Theory of turbulent plasma.} Springer (1977)
\bibitem{I91}
S. Ichimaru, \textit{Statistical Plasma Physics}, Westview (1991)


\bibitem{OYS07}
N. Okamoto, K. Yoshimatsu, K. Schneider et al., {\it Phys. Fluids} {\bf 19} (2007), 11509
\bibitem{ZSIG07}
K.P. Zybin, V.A. Sirota, A.S. Il'in, A.V, Gurevich, {\it JETP} {\bf 105} (2007), 455
\bibitem{ZSIG08}
K.P. Zybin, V.A. Sirota, A.S. Il'in, A.V, Gurevich, {\it Phys. Rev. Lett.} {\bf 100} (2008), 174504
\bibitem{ZS10}
K.P. Zybin, V.A. Sirota, {\it Phys. Rev. Lett.} {\bf 104} (2010), 154501
\bibitem{ZSI10}
K.P. Zybin, V.A. Sirota, A.S. Il'in, {\it Phys. Rev.} {\bf E82} (2010), 056324
\bibitem{ZS12}
K.P. Zybin, V.A. Sirota, \textit{Longitudinal and transverse velocity scaling exponents from merging of the vortex filament and multifractal
models}, arXiv:1204.1465


\bibitem{K02}
J.A. Krommes, {\it Phys. Reports} {\bf 360} (2002), 1


\bibitem{KLM13}
M. Kirakosyan, A. Leonidov, B. M\"{u}ller, in preparation


\bibitem{T72}
V.V. Tamoykin, {\it Astrophysics and Space Science}\; {\bf 16} (1972), 120
\bibitem{KL08a}
M.R. Kirakosyan, A.V. Leonidov, "Stochastic Jet Quenching in High Energy Nuclear Collisions", arXiv:0810.5442 [hep-ph]
\bibitem{KL08b}
M.R. Kirakosyan, A.V. Leonidov, "Energy Loss in Stochastic Abelian Medium", Proc. Quarks 2008, Zagorsk, Russia, arXiv:0809.2179 [hep-ph]

\bibitem{Dup66}
T. Dupree, {\it Phys. Fluids} {\bf 9} (1966), 1773
\bibitem{AMY2005}
P. Arnold, G. D. Moore, L. G. Yaffe, {\it Phys. Rev.} {\bf D72} (2005) 054003
















\bibitem{MMW2007}
A. Majumder, B. M\"{u}ller, X.N. Wang, {\it Phys. Rev. Lett.}\; {\bf 99} (2007), 192301


\bibitem{KLM2012a}
M. Kirakosyan, A. Leonidov, B. M\"{u}ller, \textit{Turbulence-Induced Instabilities in EP and QGP}, arXiv:1212.6555 [nucl-th]

\bibitem{KLM2012b}
M. Kirakosyan, A. Leonidov, B. M\"{u}ller, in preparation




\end{thebibliography}
\end{document}